\documentclass[11pt]{article}

\usepackage{graphicx,stfloats}
\usepackage{color,xcolor,colortbl}
\usepackage{epsfig}
\usepackage{psfrag}
\usepackage{subfigmat}
\usepackage{subeqnarray}
\usepackage[colorlinks]{hyperref}
\usepackage{amsmath,amsfonts,amsthm}
\usepackage[mathscr]{euscript}
\usepackage{algorithm2e}
\usepackage{siunitx}
\usepackage{rotating}
\usepackage{bm,bbm}
\usepackage{siunitx}
\usepackage[version=3]{mhchem}
\usepackage{blindtext}
\usepackage{float}
\usepackage{authblk}
\usepackage[square,sort,comma,numbers,compress]{natbib}

\newcommand\numberthis{\addtocounter{equation}{1}\tag{\theequation}}

\graphicspath{{./figure/}}
\makeatletter
\newcommand\rlarrows{\mathop{\operator@font \rightleftarrows}\nolimits}
\makeatother
 \def\0vec{{\mbox{\boldmath$0$}}}

\title{Application of Pareto-efficient combustion modeling framework to large eddy simulations of turbulent reacting flows}

\author[1]{Hao Wu\thanks{wuhao@stanford.edu}}
\author[1]{Peter C. Ma\thanks{peterma@stanford.edu}}
\author[2]{Thomas Jaravel\thanks{tjaravel@stanford.edu}}
\author[1,2]{Matthias Ihme\thanks{mihme@stanford.edu}}
\affil[1]{Department of Mechanical Engineering, Stanford University, 
			\protect\\Stanford, CA 94305, United States}
\affil[2]{Center for Turbulence Research, Stanford University, 
			\protect\\Stanford, CA 94305, United States}

\begin{document}

\maketitle

\begin{abstract}
In the application of the combustion models based on low-dimensional manifolds (for instance flamelet models) to large-eddy simulation (LES) of reactive turbulent flows, the modeling simplifications of the combustion process is a critical source of uncertainty in addition to those due to the turbulent closure model and numerical discretization. The ability to quantitatively assess this uncertainty in absence of the reference result is vital to the reliable and predictive simulations of practical combustion devices.In the present study, the Pareto-efficient combustion (PEC) framework is extended to adaptive LES combustion simulations of turbulent flames. The key component of the PEC framework is the so-called manifold drift term. Its extension LES is proposed to make such assessment by examining the compliance of a particular combustion model in describing a quantity of interest with respect to the underlying flow field representation. With the focus on improving predictions of $\ce{CO}$ emissions of flamelet-based combustion models, this work employs PEC to augment the flamelet/progress variable (FPV) model through local sub-model assignment of the finite-rate chemistry (FRC) model. To this end, a series of LES-PEC calculations are performed on a piloted partially-premixed dimethyl ether flame (DME-D), using a combination of FPV and FRC models. The drift term is utilized in the PEC framework to the estimate the model related error for quantities of interest. The PEC approach is demonstrated to be capable of significantly improving the prediction of $\ce{CO}$ emissions compared with the monolithic FPV simulation. The improved accuracy is achieved by enriching the FPV model with FRC in regions where the lower-order model is determined insufficient through the evaluation of drift terms.
\end{abstract}




\section{Introduction}
\label{sec_introduction}

The modeling of turbulent combustion is complex and requires the consideration of different physico-chemical processes involving a vast range of time and length scales as well as a large number of scalar quantities. Consequently, requirements for computational resources to perform detailed simulations that `directly' capture the oxidation of realistic fuels remains intractable in practical applications. To reduce the computational complexity, various combustion models are developed. Many of them can be abstracted using a lower-dimensional manifold representation~\cite{POPE_PCI2013}. Common to these techniques is the representation of the thermo-chemical state space in terms of a reduced set of scalars, whose evolution can be computed at a reduced cost.

Based on the manifold abstraction, the Pareto-efficient combustion (PEC) framework~\cite{WU_SEE_WANG_IHME_CF2015,wu2015fidelity,seeadaptive} was developed to enable the adaptive utilization of combustion models for reacting flow simulations by using a collection of combustion models through sub-model assignment in a single simulation and making informed choice of the locally appropriate model. PEC is formulated to require minimal user input, consisting of (i) a set of candidate combustion models that can be described by the manifold abstraction, (ii) a set of quantities of interest (QoI) such as emissions of $\ce{CO}$ denoted by $Q$, and (iii) a weight coefficient $\lambda$ to balance computational cost and desirable model accuracy. Using these information, the combustion sub-model assignment is dynamically determined by weighing the estimated local error and cost among the candidate models, while ensuring essential conservation properties and smooth transition between sub-models.

One common type of manifold models is the class of flamelet-based models, in which the manifolds are constructed by using canonical flame configurations of low spatial dimensionality. These models are parameterized by few manifold-describing parameters and require explicit tabulation. Models of this class have been widely adopted for large-scale reacting flow simulations due to their computational efficiency and versatility. Despite the popularity of these reduced-order models, they are often considered inadequate in the accurate prediction of emissions (e.g. $\ce{CO}$, $\ce{NO}$) for complicated multi-regime applications~\cite{IHME_SEE_CF2010,GOKTOLGA_UGUR_OIJEN_GOEY,lamouroux2014tabulated,ma2017flamelet,wu2015modeling,esclapez2017fuel}, as these species are particularly sensitive to local flow field and non-equilibrium effects.

\begin{figure}[!htb!]
\centering
\includegraphics[width=0.6\textwidth]{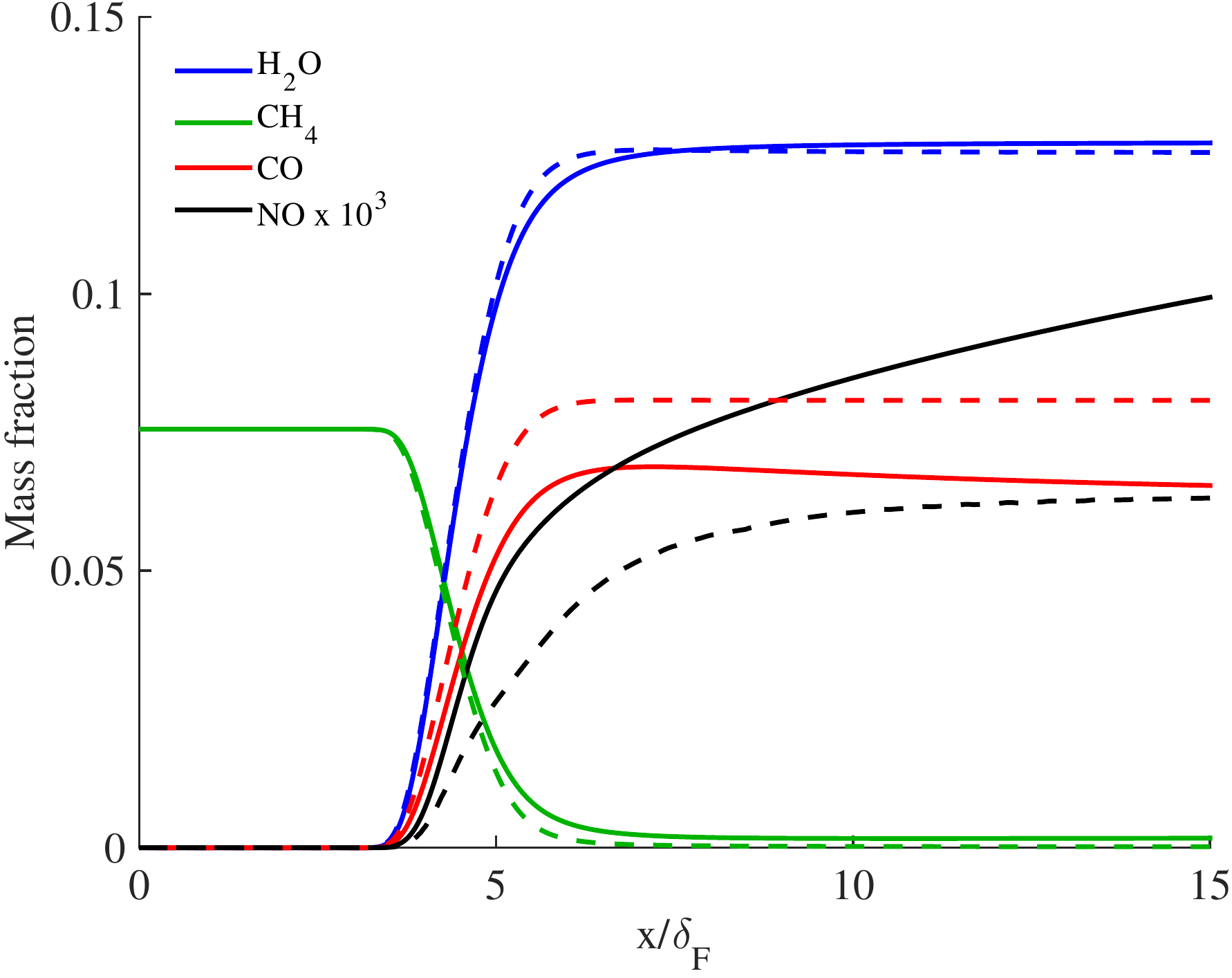}
\caption{Comparisons of $\ce{H_2O}$, $\ce{CH_4}$, $\ce{CO}$, and $\ce{NO}$ mass fractions between a freely propagating premixed flame (solid) and a moderately stratified flames (dashed). The initial conditions are methane/air mixture at ambient condition and the equivalence ratio of $\phi = 1.4$ for both cases. \label{fig:triple_flame}
}
\end{figure}

For instance, due to the reduced computational complexity, flamelet-type models are often employed in simulations of practical combustion systems such as gas turbines, internal combustion engines, rocket motors, and furnaces. Despite the presence of mixed and multiple combustion regimes in these systems, reasonable accuracy of important flow-field quantities, such as major species and heat release, have been obtained. This seemingly contradictory result can be demonstrated by a test case shown in Fig.~\ref{fig:triple_flame}. Two flames of the same equivalence ratio ($\phi=1.4$) are considered here: a 1D freely propagating premixed flame (dashed lines) and a moderately stratified flame (solid lines). The detailed configuration of the stratified flame is discussed in~\cite{WU_SEE_WANG_IHME_CF2015}. Given the very good agreement in the mass faction profiles of $\ce{H_2O}$ and $\ce{CH_4}$, the stratified flame can be considered as being ``adequately" described by the freely propagating flame in the premixed regime as far as these major species are concerned. However, the large deviations of minor species mass fractions such as $\ce{CO}$ and $\ce{NO}$ indicate that the deviation from the premixed asymptote is highly significant, such that the stratified flame cannot be modeled by a purely premixed flame. Therefore, coarse-grained regime indicators that are solely based on major species mass fractions or progress variable are insufficient to guide model selection for high-fidelity simulations. Furthermore, this result also suggests that the generic notion of a premixed or non-premixed combustion region in a mixed-regime environment can drastically oversimplify the underlying combustion physics.

The difficulty in accessing the deficiency of a given manifold-based combustion model without the prior information of a reference solution can be addressed by the so called drift term~\cite{WU_IHME_2016}. The drift term takes a different approach to the assessment of ``model adequacy", which is to test the compatibility between the combustion model and the local flow field representation. The compatibility can be quantified by the drift term, which is based on the analysis of manifold geometry and the state-space dynamics on it~\cite{POPE_PCI2013}. Quantifying the incompatibility provides a direct assessment of the model's applicability and the species-specific error that is introduced by the manifold without prior knowledge of the true solution. In addition, this method does not rely on the underlying assumptions or how the manifold is constructed. As a result, it is universal and applicable to all models under the lower-dimensional manifold abstraction.

This is approach was developed as a key component of the PEC framework. With this, the PEC-framework can dynamically determine the local utilization of different manifold-based combustion models in simulations of chemically reacting flows. By locally enriching the flamelet models using models of higher physical fidelity, the PEC approach was demonstrated to improve the prediction of emissions while incurring lower cost compared with simulations with finite-rate chemistry uniformly applied~\cite{WU_SEE_WANG_IHME_CF2015}.

In the present study, an extension of the drift term is formulated for LES equations. The objective of this work is to extend the application of PEC to LES of turbulent flames. To facilitate cost-effective realization of local model adaptation for large-scale 3D calculations, efficient computational strategies are developed for solving finite-rate chemistry, evaluating the drift term for error estimation, and scalable parallelization. The combined framework is applied to LES of a piloted turbulent dimethyl ether (DME) jet flame using a combination of the flamelet/progress variable approach (FPV) and finite-rate chemistry (FRC). The predictive capability and computational efficiency of the PEC framework are analyzed in comparison to nonadaptive simulations using only FPV or FRC model.

\section{PEC framework}
\label{sec_pec_framework}
As discussed in Sec.~\ref{sec_introduction}, the PEC-framework aims to improve the cost and accuracy of combustion simulations by applying a spatially heterogeneous sub-model assignment that is dynamically adaptive during the LES calculation. Three key components of PEC are discussed in this section: (i) the model assignment based on the balance between efficiency and fidelity of combustion models, (ii) the drift term used to estimate the model error without prior information of a reference solution, and (iii) the coupling of scalar transport equations among different models. The interested readers are referred to~\cite{WU_SEE_WANG_IHME_CF2015} for a detailed description of the PEC formulation.

\subsection{Governing equations}
The system of Favre-filtered fully compressible Navier-Stokes equations is considered in this work. The filtered conservation laws of mass, momentum, total energy, and reactive scalars takes the following form:
\small
\begin{subequations}
    \label{eqn_les}
    \begin{align}
        {\partial_t \overline{\rho}} +\nabla \cdot (\overline{\rho} \widetilde{\boldsymbol{u}}) &= 0\,,\\
        {\partial_t (\overline{\rho} \widetilde{\boldsymbol{u}})} + \nabla \cdot (\overline{\rho} \widetilde{\boldsymbol{u}} \widetilde{\boldsymbol{u}} + \overline{p} \boldsymbol{I}) &= \nabla \cdot \overline{\boldsymbol{\tau}}_{v + t} \,,\\
        {\partial_t  (\overline{\rho} \widetilde{e})} + \nabla \cdot [\widetilde{\boldsymbol{u}}(\overline{\rho} \widetilde{e} + \overline{p})] &= \nabla \cdot (\overline{\boldsymbol{\tau}}_{v + t} \cdot \widetilde{\boldsymbol{u}}) -\nabla \cdot \overline{\boldsymbol{q}}_{v + t} \,,\\
        \partial_t (\overline{\rho} \widetilde{\boldsymbol{\phi}})
	\label{eqn_scalar_les}
    + \nabla \cdot (\overline{\rho} \widetilde{\mathbf{u}} \widetilde{\boldsymbol{\phi}}) & =
    	-\nabla \cdot \widetilde{\boldsymbol{j}}_{\text{v+t}} + \widetilde{\boldsymbol{S}}_{{{\phi}}}, 
    \end{align}
\end{subequations}
\normalsize
where $\boldsymbol{\tau}$ is the stress tensor, $\boldsymbol{q}$ is the heat flux, and $p$ is the pressure obtained from the equation of state.

For the scalar equation in Eq.~\ref{eqn_scalar_les}, $\widetilde{\boldsymbol{\phi}}$ is defined to be the concatenation of manifold-describing variables of candidate models: $
	\widetilde{\boldsymbol{\phi}} = 
    	\begin{bmatrix} 
        	\widetilde{\boldsymbol{\phi}}_{m_1} ,\, \cdots ,\, \widetilde{\boldsymbol{\phi}}_{m_N}
        \end{bmatrix} ^ T
$, with the set of candidate models being $M = \{ m_1 ,\, \cdots ,\, m_N \}$. The manifold-describing variables are used to evaluated the filtered chemical source term denoted by $\widetilde{\boldsymbol{S}}_{{{\phi}}}$ as well as the reconstructed species mass fractions. The sum of fluxes due to molecular diffusion and subgrid-scale (SGS) transport is denoted by $\widetilde{\boldsymbol{j}}_{\text{v+t}}$. 

Pressure, temperature, and thermo-physical properties are obtained from the density, internal energy, and the species mass fractions reconstructed from the chosen manifold. The equation of state and thermodynamics are modeled consistently over the entire domain, independent of the manifold adaptation. 

\subsection{Pareto-efficient model assignment}
\label{sec_pec_model_assgn}
The model assignment, which is represented by the mapping $\mathcal{M}: \Omega \to M$ , determines the model of choice at any point in the computational domain $\Omega$. The assignment is obtained in the PEC framework by minimizing the weighted sum of estimated error and cost:
\begin{equation}
\label{eqn_opt}
 \underset{\mathcal{M}: \Omega\to {M}}\min \bigg(
 	\underbrace{\int_{\Omega} e(\mathcal{M}(\mathbf{x}), \mathbf{x}) d\mathbf{x}}_{\text{estimated error}} + 
	\lambda \underbrace{\int_{\Omega} c(\mathcal{M}(\mathbf{x}), \mathbf{x}) d\mathbf{x}}_{\text{estimated cost}}
	\bigg)\,.
\end{equation}
The coefficient $\lambda$ is a user-input parameter, expressing the preference in the balance between efficiency and fidelity. As such, model assignment of lower cost but higher error will be chosen with an increased value of $\lambda$. The estimated local cost ($c$) is a model-specific constant, which for the present case is chosen to be equal to the number of transported scalars of the combustion model. The local error ($e$) is defined as the weighted sum of error estimated for the QoIs: 
\begin{equation}
	e(m, \mathbf{x}) = \sum_{\psi \in Q} \left|\frac{\mathcal{D}^{m}_{\psi}}{B_{\psi}} \right| \, ,
\end{equation}
where $m = \mathcal{M}(\mathbf{x})$ is the locally applied combustion model; $\psi$ is a quantity included in the set of QoIs, denoted by $Q$: and $B_{\psi}$ is the scaling constant. The formulation of the drift term, $\mathcal{D}^{m}_{\psi}$, will be described in the follow section. A greedy strategy is employed to approximate the solution to Eq.~\ref{eqn_opt} and the detailed algorithm is discussed in~\cite{WU_SEE_WANG_IHME_CF2015}. 

\subsection{Drift term}
\label{sec_drift_term}

\begin{figure}[!htb!]
\centering
\includegraphics[width=0.8\textwidth]{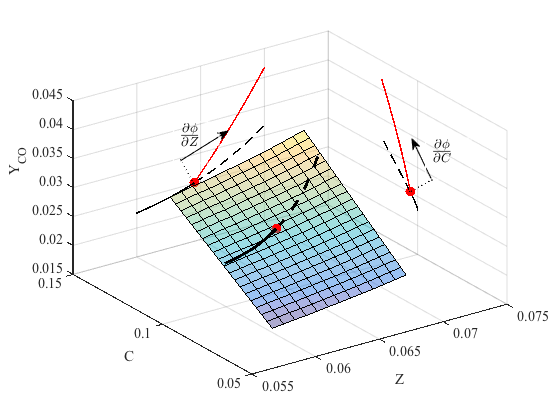}
\caption{Drift from flamelet-based manifold for $\phi = Y_{\ce{CO}}$. The trajectory of the solid black line represents the path of $\widehat{Y}_{\ce{CO}}$ determined by flamelet-based combustion model. The red line represents the trajectory of ${Y}_{\ce{CO}}$ if the full system of equations were solved at the point marked by the red dot. The drift term calculates the initial growth rate of departure between the red line and the dashed black line, marking the path of $\widehat{Y}_{\ce{CO}}$ if the flamelet-based manifold were still in use. \label{fig:man_drift_illus}
}
\end{figure}

The drift term for manifold-based combustion models is the QoI-specific error estimator that takes into account the information of the local combustion model and the instantaneous profiles of scalars and hydrodynamic quantities. It calculates the initial growth-rate of the error if the solution were initialized from a given manifold model. Therefore, it identifies the error induced by the incompatibility between the local flow field and the structure of the combustion model~\cite{WU_IHME_2016}. For a given combustion model $m$, the drift term for the QoI $\psi \in Q$ is defined as
\begin{equation}
\label{eqn_drift}
	\mathcal{D}^{m}_{\psi}  = \left.D_t \big( \overline{\rho} \widetilde{\psi} \big) \right|_{\widetilde{\psi}=\widetilde{\psi}^*_{m}} - D_t \big( \overline{\rho} \widetilde{\psi}^*_{m} \big) \,,
\end{equation}
where $D_t$ denotes the substantial derivative and $\psi^*_{m}$ is the QoI determined from the manifold of model $m$. The first term ($\left.D_t \big( \overline{\rho} \widetilde{\psi} \big) \right|_{\widetilde{\psi}=\widetilde{\psi}^*_{m}}$) represents the evolution of $\psi$ following the transport equations, if it were initialized by the manifold determined value $\psi^*_{m}$. The second term ($D_t \big( \overline{\rho} \widetilde{\psi}^*_{m} \big)$) represents the change of $\psi^*_{m}$ following the evolution of the manifold-describing variables, which can be obtained via the chain rule as~\cite{WU_IHME_2016}
\begin{equation}
\label{eqn_drift_jac}
	D_t \big( \overline{\rho} \widetilde{\psi}^*_{m} \big)= \frac{\partial \widetilde{\psi}^*_{m} }{\partial \widetilde{\boldsymbol{\phi}}_m}  \cdot D_t \big( \overline{\rho} \widetilde{\boldsymbol{\phi}}_m \big) + \widetilde{\psi}^*_{m} D_t \overline{\rho}\,.
\end{equation}
An illustration of this concept is shown in Fig.~\ref{fig:man_drift_illus}. The flamelet-based lower-dimensional manifold has the reaction coordinate being the mixture fraction $Z$ and the progress variable $C$, thus $\boldsymbol\psi = \big[Z,\, C \big]^T$. The trajectory of the solid black line represents the dynamics of the modeled quantity in the state space. The path of the $\ce{CO}$ mass fractions is confined to the 2D surface due to the manifold constraint. The red dot represents the time instance, at which the drift term is evaluated. The red trajectory represents the future path of $Y_{\ce{CO}}$ if the full system of equations were solved from now on using the detailed model. Depending on the local flow field conditions, the path can leave the 2D manifold and depart from the dashed black line, marking the path of $Y_{\ce{CO}}$ if the flamelet-based manifold were still in use. 

In the present study, the QoIs are chosen to be the mass fractions of selected species. Therefore, all the terms present in Eq.~\ref{eqn_drift} can be obtained directly from the simulation using model $m$. As such, the drift term is capable of determining the applicability of a manifold-based combustion model at the absence of prior information from a reference solution. 

\subsection{Coupling between combustion models}
The scalar equations in Eq.~\ref{eqn_scalar_les} are selectively solved based on the dynamically determined model assignment $\mathcal{M}$. As a result, the transport equation of $\widetilde{\boldsymbol{\phi}}_{m}$ is solved where $\mathcal{M}(\mathbf{x}) = m$.
Coupling between scalar transport equations among sub-domains using different models is achieved via the mechanism of imputing the deactivated manifold-describing variables using those of the activated model. To ensure consistency at the interface of sub-models, the following reconstruction operation is applied before the right-hand-side evaluation of the spatially-discretized governing equations: 
\begin{equation}
\label{eqn_recon}
    \widetilde{\boldsymbol{\phi}}_{m^\prime} (\mathbf{x}) = \widetilde{\boldsymbol{\phi}}^R_{m^\prime} (\widetilde{\boldsymbol{\phi}}_{m}(\mathbf{x})),\, \text{where } \mathcal{M}(\mathbf{x}) = m,
\end{equation}
for all deactivated models $m^\prime \ne \mathcal{M}(\mathbf{x})$. For the FPV and FRC models used in this study, the specific form of the reconstruction operation $\widetilde{\boldsymbol{\phi}}^{R}$ is described in Sec.~\ref{sec_pec_fpv_frc}.

The essential conservation properties for mass, momentum, and total energy are preserved during the adaptation procedure as the corresponding conservation equations are universal among sub-domains. The smoothness of the scalar fields is ensured by the combination of the reconstruction operation and the drift term which detects sharp transitions through the entailed diffusion operator.

\section{Computational approach}
\label{sec_comp_approch}

\subsection{Spatial and temporal discretization}
\label{sec_num_method}
LES with PEC-based combustion model adaptation is performed using an unstructured finite-volume solver, CharLES$^x$~\cite{KHALIGHI_NICHOLS_HAM_LELE_MOIN_AIAA2011,wu2017mvp,ma2017entropy,lv2017underresolved}, developed at CTR. The fully compressible multi-species Navier-Stokes equations are spatially-discretized using a modified hybrid central-WENO scheme~\cite{LARSSON_GUSTAFSSON_2008,JIANG_SHU_1996}, which is of low numerical dissipation while being stable for discontinuities across contact interface.

The convection-diffusion and reaction operators of the system are separated by a steady-state preserving $2$nd-order operator splitting scheme~\cite{wu2017application,WU_MA_IHME_2017}. The time-stepping of the convection-diffusion part is carried out via a 3rd-order Runge-Kutta (SSP-RK3) method~\cite{GOTTLIEB_SHU_TADMORE_2001}. The spatially independent systems of chemical reactions are integrated using a semi-implicit $4$th-order Rosenbrock-Krylov (ROK4E) scheme~\cite{TRANQUILLI_SANDU_2014,WU_IHME_ROK4E_2016}. The stability of the ROK4E scheme is achieved through the approximation of the Jacobian matrix by its low-rank Krylov-subspace projection. As few as three right-hand-side evaluations being performed over four stages. Overall, a minimum of $3.5$ source-term evaluations is required per convection-diffusion step, the cost of which is comparable to $3$ evaluations required for a fully-explicit scheme. 

\subsection{Parallelization and load balancing}
\label{sec_scal_paral}
For parallel computation on distributed architectures, the unstructured mesh is partitioned into multiple domains. The heterogeneity in the memory layout is accounted for by using a two-layer partitioning strategy \cite{BERMEJO_BODART_LARSSON_2013}. The partition is first conducted at the inter-node level followed by a second partitioning step at the intra-node level. The cost at each control-volume is the sum of the local combustion modeling cost discussed in Sec.~\ref{sec_pec_model_assgn} and the fixed cost of solving the compressible Navier-Stokes equations. In addition to the cost-aware domain decomposition, another level of load balancing is performed for the time-stepping of chemical reactions where the FRC model is used. The computational task of the time integration is redistributed among the processors by exploiting the spatial locality of chemical reactions. The redistribution is conducted via a parallelized rebalancing scheme~\cite{WU_MA_IHME_2017} based on the algorithm of Aggarwal et al.~\cite{AGGARWAL_MOTWANI_ZHU_2003}. The communication overhead is justified by the arithmetic intensity associated with the integration of combustion chemistry.

\section{Experimental and computational setup}
\subsection{Experimental configuration}
In the present work, PEC is applied to LES of a turbulent dimethyl ether (DME) jet flame, which was derived from the canonical Sydney/Sandia piloted jet flame series and was experimentally investigated at Sandia National Laboratories~\cite{FUEST_BARLOW_CHEN_DREIZLER,FUEST_MAGNOTTI_BARLOW_SUTTON_2015}. The central fuel jet consists of a mixture of DME and air with an equivalence ratio of $\phi = 3.6$, resulting in a stoichiometric mixture fraction of $Z_{st} = 0.353$. The pilot stream consists of a burned mixture of $\ce{C_2H_2}$, $\ce{H_2}$, $\ce{CO_2}$, and $\ce{N_2}$, which is chosen to match a premixed DME flame with $\phi = 0.6$. The fuel jet diameter is $D = 7.45$ mm. The pilot annulus has inner and outer diameters of $8$ mm and $18.2$ mm. For this study, the DME-D condition is chosen. The bulk jet velocity is $U_{\text{bulk}} = 49.5$ m/s, which corresponds to a Reynolds number of $\text{Re} = 29,300$. The unburned velocity of the pilot stream is $1.1$ m/s and the velocity of the air co-flow is $0.9$ m/s. 

\begin{figure}[bt]
\centering
\includegraphics[trim={0.2cm 9.75cm 0.1cm 8cm},clip,width=0.8\textwidth]{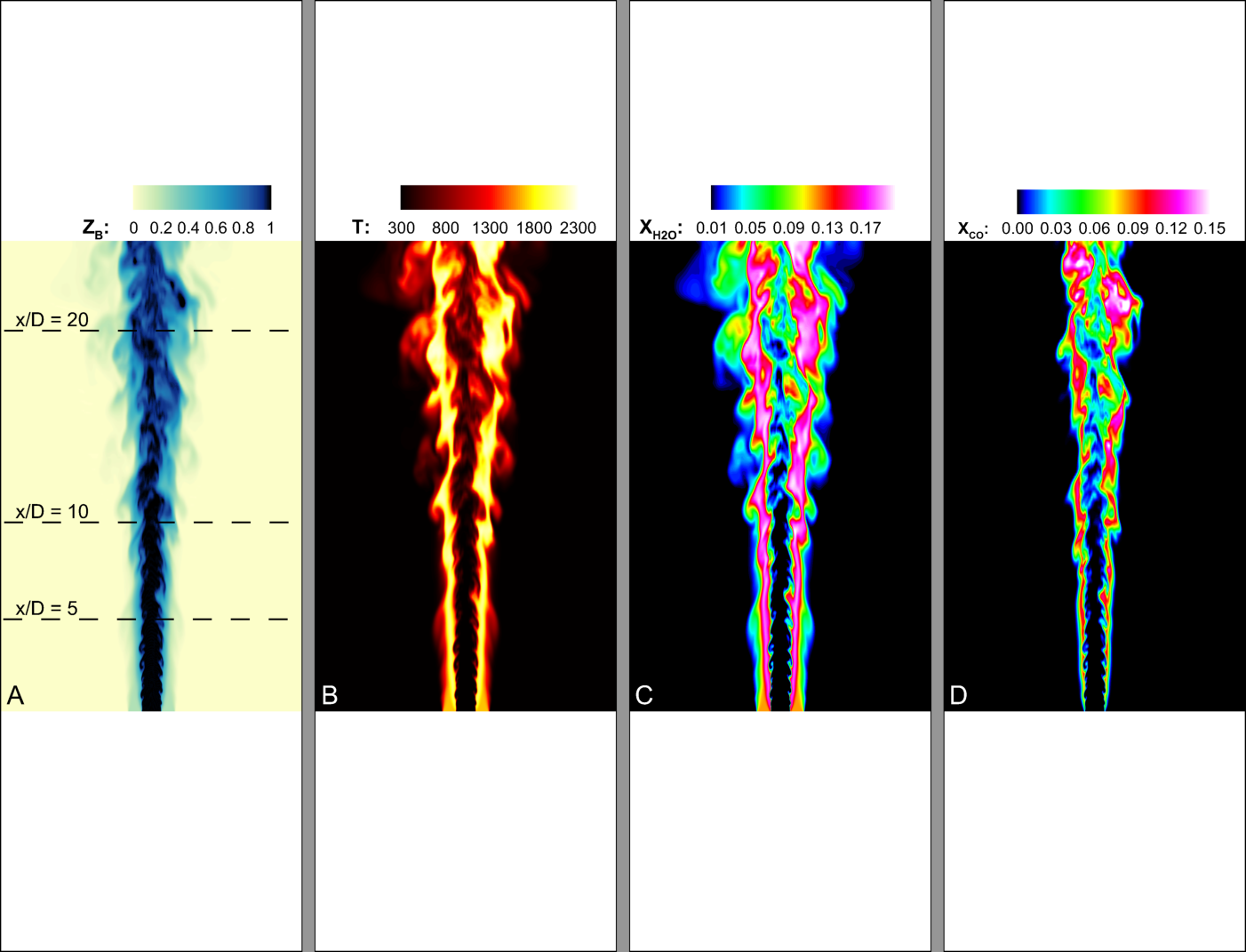}
\caption{Instantaneous fields from LES-FRC. 
		 (A) Mixture fraction, 
         (B) Temperature (K), 
         (C) $\ce{H_2O}$ mole fraction, 
         (D) $\ce{CO}$ mole fraction.
\label{fig_FRC_INST}}
\end{figure}

\subsection{Numerical setup}
The LES is performed on an unstructured mesh with $10.3$ million hexahedral elements. The minimal grid spacing at the jet outlet is $0.1$ mm. The velocity inflow profile of the fuel stream is generated using digital filtering~\cite{KLEIN_SADIKI_JANICKA_JCP2003}, with mean and turbulent quantities matching the experimental data. The nozzle wall is considered as non-slip and adiabatic. 

For both the FPV and FRC models, the combustion chemistry of the DME-air flame is represented using an 18-species kinetic scheme that was reduced from the 58-species mechanism of Zhao et al.~\cite{ZHAO_CHAOS_KAZAKOV_DRYER_2008}, using a combination of skeletal reduction and linear-QSSA~\cite{LU_LAW_2006} method.           The detailed description and testing for the reduced mechanism is provided as supplementary material. The molecular diffusion is modeled with the species-specific non-unity Lewis numbers assumed constant throughout the flame. For mixture fraction and progress variable, the Lewis numbers are set to be unity. The Lewis numbers for the species are calculated at the equilibrium condition of a stoichiometric mixture of DME and air. The Vreman model~\cite{VREMAN_POF2004} is used for turbulent subgrid stresses. The subgrid-scale turbulent-chemistry interaction is accounted for by using the dynamic thickened-flame model~\cite{legier2000dynamically}, with a maximal thickening factor of $3$. The same closure model is applied to both FPV and FRC models. 

When applied to the $18$-species reduced DME mechanism, the efficient integration of chemical source terms accounts for less than $20\%$ of the overall computational cost in the nonadaptive FRC simulation, while the transport equations of the species mass fractions account for $65\%$. Though both components of cost can be reduced via the limited usage of FRC in PEC simulations, reducing the number of transport equations via PEC is critical to meaningful cost saving. For more complex mechanisms, the cost associated with chemical reactions in FRC can be further reduced via additional adaptation on the chemical mechanism~\cite{xie2017dynamic,yang2017parallel,yang2017global,yang2017comparison,yang2016parallel}. Though not considered in this study, the adaptation of kinetic models, as demonstrated in~\cite{WU_SEE_WANG_IHME_CF2015}, can also be incorporated in the PEC framework thanks to the universality of the manifold concept.

\begin{figure}[!hbt]
\centering
\includegraphics[trim={0.0cm 0.3cm 0.0cm 0.3cm},clip,width=0.80\linewidth]{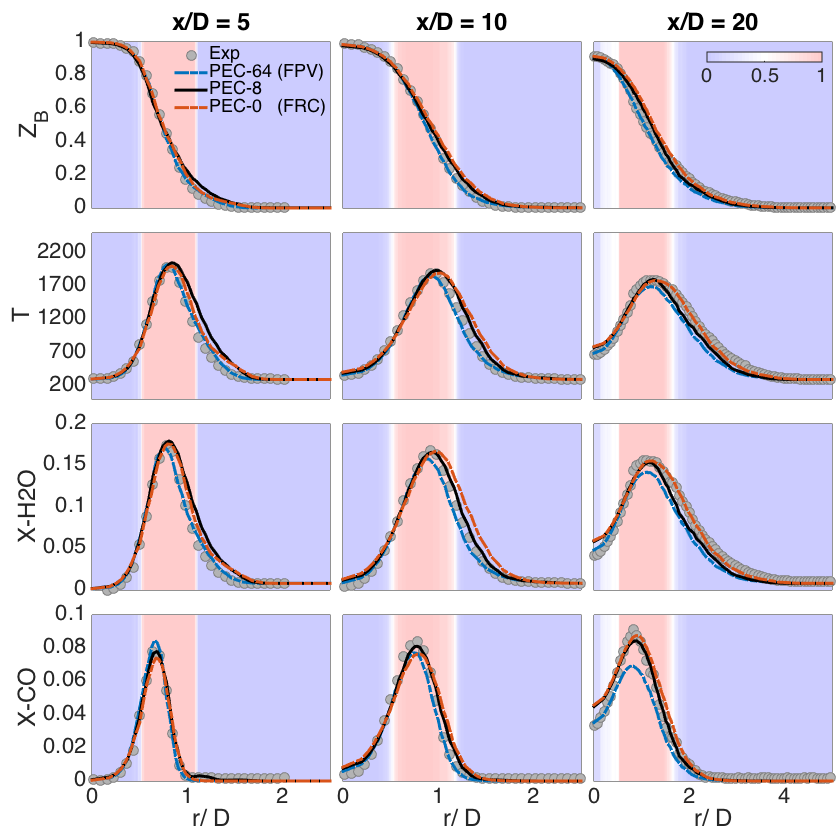}\\[-1em]\setlength{\belowcaptionskip}{-8pt} 
\caption{Comparison of the experimental and computed radial profiles of the mean Bilger mixture fraction, temperature (K), mole fractions of $\ce{H_2O}$ and $\ce{CO}$, from PEC-64 ($\lambda = 0.64$, FPV), PEC-8 ($\lambda = 0.08$), and PEC-0 ($\lambda = 0.00$, FRC), at the axial positions of $x/D =\{5, \, 10, \, 20\}$. The probability of choosing FRC over FPV in the case of PEC-8 is color coded.
\label{fig_RADIAL_COMP}}
\end{figure}

\subsection{Pareto-efficient combustion model}
\label{sec_pec_fpv_frc}
The candidate models considered in this study are the flamelet/progress variable (FPV) and the finite-rate chemistry (FRC) models. The FPV model is chosen since DME jet flame is largely diffusion-controlled. More complex flames could benefit from using additional sub-models. As such, the manifold-describing variables are defined as $
	\widetilde{\boldsymbol{\phi}} = 
    	\begin{bmatrix} 
        	\widetilde{\boldsymbol{\phi}}_{FPV} ,\, \widetilde{\boldsymbol{\phi}}_{FRC}
        \end{bmatrix} ^ T$.
        
The choice of using the dynamic thickened-flame model  for both FRC and FPV allows the difference observed in the prediction of QoIs to be analyzed without the potential complication from the variation of closure models. With the flame being artificially thickened, the FPV manifold is described by the filtered mixture fraction and progress variable, $\widetilde{\boldsymbol{\phi}}_{FPV} = \begin{bmatrix} \widetilde{Z} ,\, \widetilde{C} \end{bmatrix} ^T$, which we note is different from the conventional practice of using a presumed-PDF. For the FRC model, $\widetilde{\boldsymbol{\phi}}_{FRC} = \begin{bmatrix} \widetilde{Y}_1 ,\, \cdots ,\, \widetilde{Y}_{N_S-1} \end{bmatrix} ^T$, leading to an identity mapping between the manifold-describing variables and the filtered species mass fractions. The Jaocian-vector multiplication can be efficiently calculated by finite-difference approximation:
\begin{align*}
\label{eqn_drift_fd}
& \frac{\partial \widetilde{Y}^*_k }{\partial \widetilde{Z}}  \frac{D}{Dt} \overline{\rho} \widetilde{Z} + \frac{\partial \widetilde{Y}^*_k }{\partial \widetilde{C}} \frac{D}{Dt} \overline{\rho} \widetilde{C} 
\approx \\
& \quad \bigg[
	\widetilde{Y}^*_k(\widetilde{Z} + \epsilon \frac{D}{Dt} \overline{\rho} \widetilde{Z}, \widetilde{C} + \epsilon \frac{D}{Dt} \overline{\rho} \widetilde{C}) - \widetilde{Y}^*_k(\widetilde{Z}, \widetilde{C})
    \bigg] / {\epsilon} \, , \numberthis
\end{align*}
which is $1$st-order accurate with one additional retrieval of the tabulated values.

The progress variable for the FPV model is defined as the sum of mass fractions of $\ce{H_2}$, $\ce{H_2O}$, $\ce{CO}$, and $\ce{CO_2}$. Hence, the reconstructed progress variable for FPV, used in Eq.~\ref{eqn_recon}, is obtained from $\boldsymbol{\phi}_{FRC}$ as $\widetilde{C}^R = \widetilde{Y}_{\ce{H_2}} + \widetilde{Y}_{\ce{H_2O}} + \widetilde{Y}_{\ce{CO}} + \widetilde{Y}_{\ce{CO_2}}$, while the species mass-fraction for FRC is reconstructed via table reading: $\widetilde{Y}^R_{k} = \widetilde{Y}^*_{FPV,\,k}(\widetilde{Z}, \, \widetilde{C})$, where $\widetilde{Y}^*_{FPV,\,k}$ is the tabulated species mass fraction from FPV. The scaling constant $B_{\psi}$, shown in Eq.~\ref{eqn_drift}, is chosen to be the corresponding species production rates at the equilibrium condition of a stoichiometric DME-air mixture. Due to the consideration of preferential diffusion in the FRC model, the mixture fraction with unity Lewis number cannot be reconstructed from species mass fractions and thus the consistent transport equation for $\widetilde{Z}$ is solved throughout the domain. 

\begin{figure}[hbt]
\centering
\includegraphics[trim={0.0cm 0.0cm 0.0cm 0.0cm},clip,width=90mm]{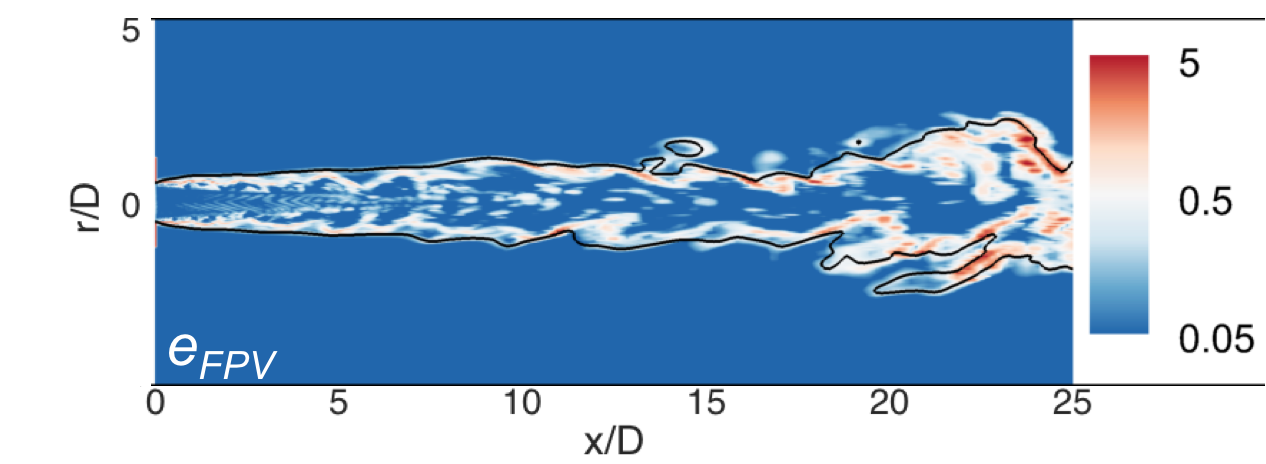}\\[-1.0em]\setlength{\belowcaptionskip}{-8pt} 
\setlength{\belowcaptionskip}{-8pt} 
\caption{Instantaneous field of local error estimation for PEC-64 (FPV), defined as the sum of the normalized drift terms for $\ce{H_2}$, $\ce{H_2O}$, $\ce{CO}$, and $\ce{CO_2}$, with stoichiometric contour in black line.
\label{fig_PEC64_INST}}
\end{figure}

\begin{figure}[ht]
\centering
\includegraphics[width=0.60\linewidth]{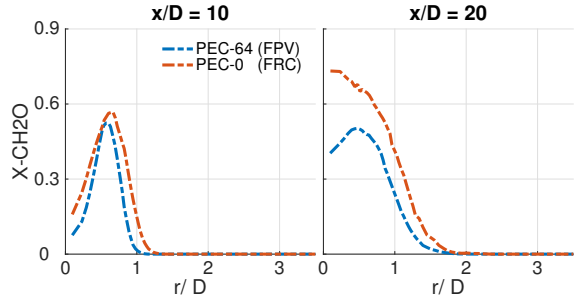}
\caption{Comparison of the computed radial profiles of the mean $\ce{CH_2O}$ mole fraction from LES-FPV and LES-FRC, at the axial positions of $x/D = 5, \, 10, \, 20$.
\label{fig_RADIAL_COMP_CH2O}}
\end{figure}

\begin{figure}[hbt]
\centering
\includegraphics[trim={0.2cm 9.75cm 0.1cm 8cm},clip,width=0.8\textwidth]{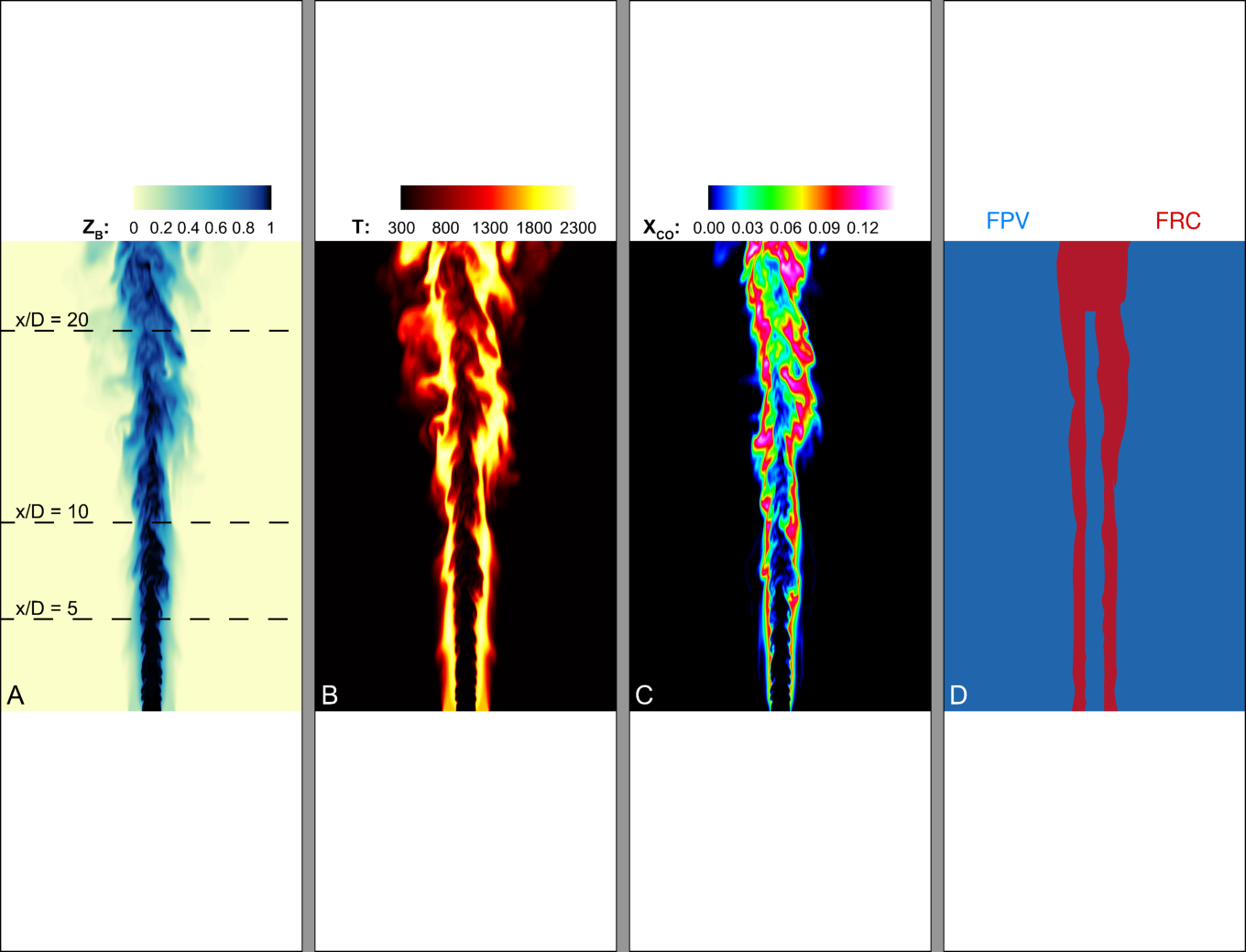}
\caption{Instantaneous fields of PEC-8. 
		 (A) Mixture fraction, 
         (B) Temperature (K), 
         (C) $\ce{CO}$ mole fraction, 
         (D) model assignment, \textcolor{blue}{FPV} and \textcolor{red}{FRC} are color coded in blue and red respectively.
\label{fig_PEC8_INST}}
\end{figure}


\section{Results and discussion}

A series of PEC simulations based on FPV and FRC models are performed. Calculations are carried out over a range of $\lambda$, specifying different levels of efficiency-accuracy preference. The large value of $\lambda$ corresponds to preferring the low-order FPV model over the FRC model, which has greater computational complexity but a higher degree of fidelity. The shift toward FPV leads to a decrease in the cost of the simulation and a potential increase in the prediction error. The model assignment is dynamically updated at the beginning of each time step. The quantities of interest specified in the series of LES-PEC calculations are the constitutive species for progress variable: $Q = \{\ce{H_2},\, \ce{H_2O},\, \ce{CO},\, \ce{CO_2} \}$. 

We first discuss the calculations at the two limits of efficiency-accuracy preference, leading to monolithic combustion model assignments of FPV ($\lambda \to \infty$) and FRC model ($\lambda \to 0$). The results obtained from intermediate values of $\lambda$ are then presented, in which the FPV and FRC models are jointly utilized in the LES calculations. Aspects of dynamic model assignment and accuracy of $\ce{CO}$ emissions are addressed. 

Two baseline simulations are performed, in which FRC and FPV are used as the monolithic combustion model over the entire domain. These calculations are obtained with the penalty weight coefficient of $\lambda = 0.00$ and $\lambda = 0.64$ respectively. The choice of extreme values of $\lambda$ indicates the overwhelming emphasis of efficiency over accuracy or vice versa, and hence the monolithic model assignment. These PEC simulations are labeled as PEC-0 and PEC-64, corresponding to the monolithic utilization of FRC and FPV respectively. The PEC-64 result is further used to initialized the rest of the calculations to avoid the propagation of information from higher fidelity simulations. 

The mean profiles of the monolithic LES results are collected at $x/D = \{5,\,10,\,20 \}$. The comparison of the computed radial profiles against the experimental measurements is shown in Fig~\ref{fig_RADIAL_COMP}. Overall, the Bilger mixture fraction, temperature, and $\ce{H_2O}$ mole fraction are well predicted by both simulations. The level of agreement is comparable to previous LES calculations performed using FPV~\cite{POPP_HUNGER_HARTL_MESSIG_CORION_FRANK_FUEST_HASSE_2015}, CMC~\cite{coriton2015imaging}, and PDF~\cite{you2017effects} approaches. Noticeable deviation from the measurement, however, can be observed for the $\ce{CO}$ mole fraction of PEC-64 (FPV) at the downstream location ($x/D = 20$), as the FPV model reduces the chemical time scales of all species to that of the progress variable. The $\ce{CO}$ mole fraction is significantly improved in PEC-0 (FRC) due to the more detailed representation of chemical reactions~\cite{yang2017sensitivity}.

The instantaneous field of the local error estimation for PEC-64 is presented in Fig.~\ref{fig_PEC64_INST}. The error estimation is defined as the sum of the normalized drift terms for $\ce{H_2}$, $\ce{H_2O}$, $\ce{CO}$, and $\ce{CO_2}$. The value is high on the rich side of the flame and gradually increases along the streamwise direction, correctly marking the regions where FPV shows deficiency. As shown in Fig.~\ref{fig_RADIAL_COMP}, the error estimation agrees well with the comparison of the instantaneous and mean fields of $\ce{CO}$ mole fraction between PEC-64 (FPV) and PEC-0 (FRC).

The difference between PEC-64 and PEC-0 can be further illustrated when intermediate species such as $\ce{CH2O}$ are considered. The mean radial profiles for the mole fractions of these species are plotted in Fig.~\ref{fig_RADIAL_COMP_CH2O}. A much higher level of $\ce{CH2O}$ is predicted by PEC-0, which is of better agreement with the $\ce{CH2O}$-LIF signals experimentally obtained~\cite{POPP_HUNGER_HARTL_MESSIG_CORION_FRANK_FUEST_HASSE_2015}. The difference is likely due to the deficiency of the FPV model in capturing the unsteady pyrolysis process that converts DME to $\ce{CH2O}$ at very rich conditions~\cite{GABET_SHEN_PATTON_FUEST_SUTTON_2013}.

The PEC-8 calculation is performed with the weight parameter chosen to be $\lambda = 0.08$. An instantaneous snapshot of this case is displayed in Fig.~\ref{fig_PEC8_INST}A~-~\ref{fig_PEC8_INST}C, showing the smooth profiles of the Bilger mixture fraction, temperature, and mole fraction of $\ce{CO}$ at the presence of heterogeneous model assignments. The choice of intermediate values of $\lambda$ results in the adaptive procedure with a mixed usage of the FPV and FRC models, as shown in Fig.~\ref{fig_PEC8_INST}D. The comparison between PEC-8 results and the baseline simulations are shown in Fig~\ref{fig_RADIAL_COMP}, with the color-coded probability of FRC usage in PEC-8. For mean radial profiles that are well predicted by both PEC-64 and PEC-0, such as Bilger mixture fraction, temperature and $\ce{H_2O}$ mole fraction, the prediction by PEC-8 is expectedly of similar accuracy. The improvement of PEC-8 over PEC-64 can be most clearly observed for $\ce{CO}$ mole fraction at $x/D = 20$, which is significantly under-predicted by PEC-64. The deficiency of the FPV model in describing the evolution of $\ce{CO}$ is correctly detected by the PEC adaptation procedure, resulting in the usage of FRC (marked in red) in regions critical to $\ce{CO}$ prediction, which accounts for $30.2\%$ of the computational domain. 

\section{Conclusions}
\label{sec_conclusion}
In this work, the PEC framework is extended to simulations of turbulent reacting flows and the method is applied to LES of a piloted turbulent DME jet flame. Based on the minimal user-specific inputs: (i) candidate combustion models, (ii) quantities of interest, and (iii) a weight coefficient representing the balance between efficiency and accuracy, PEC selects dynamically the appropriate local combustion model assignment to best describe the flame under the specified constraints for accuracy and cost. The key component of the PEC framework is the so-called manifold drift term. Its extension LES is proposed to make such assessment by examining the compliance of a particular combustion model in describing a quantity of interest with respect to the underlying flow field representation. With the goal to enrich FPV for the improved prediction of $\ce{CO}$, a series of LES-PEC simulations is performed with various levels of FRC/FPV usage. The monolithic FPV simulation is shown to be deficient at predicting $\ce{CO}$ emissions in the down-stream region of the flame. The $\ce{CO}$ prediction can be significantly improved via the enrichment of FRC occurring at as few as $30\%$ of the domain. The resulting parallel calculation is significantly less expensive compared with a highly efficient nonadaptive calculation using finite-rate chemistry uniformly, while not causing noticeable degradation in the predictive accuracy of QoIs.

\section*{Acknowledgments}
\label{Acknowledgments}
Financial support through NASA Award No.~NNX15AV04A is gratefully acknowledged. Resources supporting this work were provided by the NASA High-End Computing (HEC) Program through the NASA Advanced Supercomputing (NAS) Division at Ames Research Center.



\end{document}